\documentstyle{article}
\begin{document}

\begin{center}
{\LARGE \bf Orbital periods and waveforms of dwarf novae observed by Kepler}
\vspace{0.5cm}

{\Large \bf Albert Bruch} \\ [1ex]
Laborat\'orio Nacional de Astrof\'{\i}sica \\
Rua Estados Unidos, 154, \\
37504-364 Itajub\'a - MG, Brazil
\vspace{0.5cm}

(Published in: Res.\ Notes AAS, {\bf 5}, 188 (2021))
\vspace{0.5cm}
\end{center}

\begin{abstract}
\parindent0em 
Kepler high cadence data are used to measure the orbital periods and to 
determine the orbital waveforms of five dwarf novae. A significant
improvement of the period of V1504~Cyg is achieved, while for the other
systems periods are derived which are compatible with previous determinations. 
The orbital waveforms of the short period systems V1504~Cyg, V344~Lyr and 
V516~Lyr are very nearly sinusoidal, while the
longer period dwarf nova V447~Lyr appears almost to be a twin of U~Gem.
The unusual system KIC~9202990 exhibits distinct variations of its waveform 
as a function of brightness during its outburst cycle. 
\vspace{1ex}

\end{abstract}

\section{Introduction} 

The often uninterrupted high cadence light curves of many stars generated 
by the Kepler mission represent a treasure trove for innumerable aspects of 
variable star research. Here, I present a small study of the orbital 
periods and waveforms of five dwarf novae using all available high cadence
Kepler data. The targets are the SU~UMa type dwarf novae V1504~Cyg, 
V344~Lyr and V516~Lyr, the U~Gem type star V447~Lyr, and the
system KIC~9202990 which is unusual in the sense that instead of full
fledged dwarf nova outbursts it exhibits a continuous series of low amplitude 
modulations, considered by Ramsay et al.\ (2016) as stunted outbursts. 

Based on measurements over a time base of 6 days (Thorstensen \& Taylor 1997)
measured a period $0.06951 \pm 0.00005$~d in V1504~Cyg. 
The Kepler data permit us to refine this
value by at least one order of magnitude. The periods of all other stars
were derived from their Kepler light curves 
[V344~Lyr: 0.087903~d (Osaki \& Kato 2013b); 
V516~Lyr: 0.083999~d (Kato \& Osaki 2013); 
V447~Lyr: 0.1556270~d (Ramsay et al.\ 2012); 
KIC~9202990: 0.1659404~d (Ramsay et al.\ 2016)], 
albeit using only a subset of 
the complete data (except for KIC~9202990). Thus, only an incremental 
improvement of their precision can be expected when the entire data set is
used. However, except for the outburst states of the eclipsing system V447~Lyr 
(Ramsay et al.\ 2012) and for KIC~9202990 (Ramsay et al.\ 2016) the waveform of 
the orbital variations was never explored.

\section{The orbital periods}

In order to measure the orbital periods, only quiescent phases of the dwarf 
novae were regarded. Again, KIC~9202990 is an exception as the entire light 
curve was used. As is well known from several studies of V1504~Cyg and 
V344~Lyr (e.g., Osaki \& Kato 2013ab)
sometimes positive and/or negative superhumps are observed together with 
orbital variations. Therefore, only light
curve intervals of these stars were regarded where the orbital modulations are 
clearly present and not contaminated by superhumps signals. In the faint system 
V516~Cyg, I confirm the weak presence of the periodic signal found by 
Kato \& Osaki (2013) during some quiescent intervals (never in outburst).
Only these parts of the light curves are used here.

For the final analysis I subjected the data to a period search routine
[Lomb-Scargle algorithm (Lomb 1976, Scargle 1982) for systems with 
approximately sinusoidal modulations (see below); analysis-of-variance 
(Schwarzenberg-Czerny 1989) for the others]. In V1504~Cyg I find an 
improved period of $0.069569 \pm 0.000002$ days. 
As expected, only in this case a significant improvement of the precision 
of the period could be achieved, while in all other stars the slightly 
enlarged data base yields periods which agree
with previous determinations within the error limits. 

\section{The orbital waveforms}

To find the waveforms of these variations and their amplitudes, the
data used for the period determination were first filtered
with a low pass filter (Savitzky \& Golay 1964),
effectively removing all variations below a given cut-off time scale of
5 times the orbital period. The filtered light curves were
subtracted from the original ones, eliminating
modulations on longer time scales. The results were
phase folded on the orbital period and binned in phase
intervals of 0.01 as shown in Figure~1. The total 
amplitude of the variations (regarding 
only the well expressed orbital humps in the case of 
V447~Lyr and KIC~9202990; see below), 
were transformed into magnitudes, yielding 0.124~mag (V1504~Cyg), 0.053~mag 
(V344~Lyr), 0.804~mag (V447~Lyr), 0.017~mag (V516~Lyr) and
0.037~mag (KIC~9202990). Note that these are average values for 
those time intervals when the systems exhibited clear orbital variations. 
During other epochs, the modulations can be weaker or even absent. 

\input epsf
%--------------------------------------------------------------
\begin{figure}
\parbox[]{0.1cm}{\epsfxsize=14cm\epsfbox{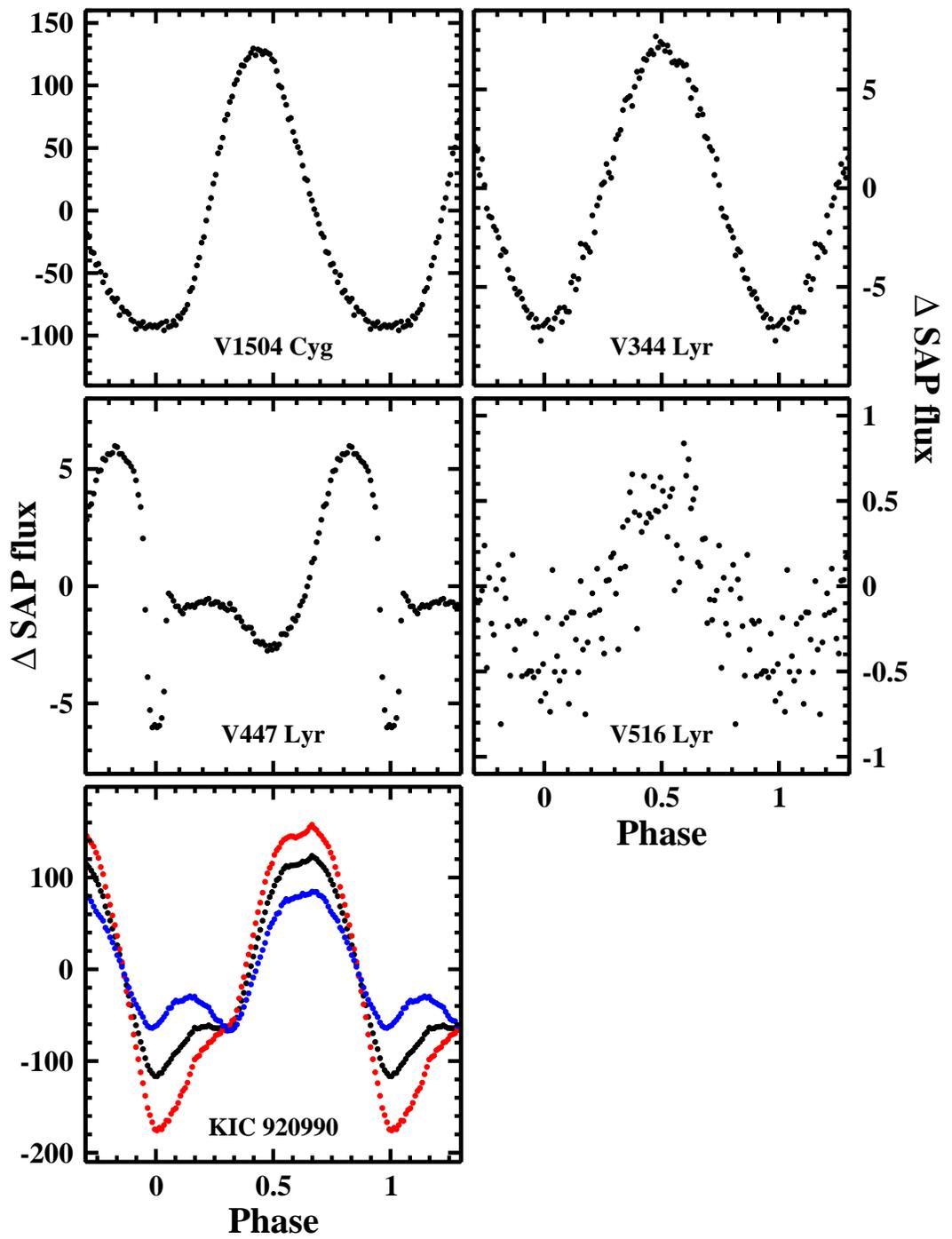}}
%      \fbox{\rule[14cm]{14cm}{0cm}}
      \caption[]{Average waveforms of the orbital variations of the dwarf 
                 novae of this study. The black graph for KIV~9202990 
                 represents the average over the entire light curve. 
                 Red and blue graphs refer to bright and the faint states, 
                 respectively.}
\label{waveform}
\end{figure}
%______________________________________________________________

The waveform of V1504~Cyg is a slightly skewed sinusoid with an indication of 
a short constant phase at minimum. Those of V344~Lyr and V516~Lyr are
almost perfect sine curves (quite noisy in V516~Lyr because of the faintness 
of the object). The orbital waveforms of the remaining two systems exhibit more 
structure and are thus more interesting.

Ramsay et al.\ (2012) found V447~Lyr to be eclipsing. They examined the waveform
during outburst, but the quiescent phase folded light curve is shown here for 
the first time. It is remarkably similar to the average
quiescent light curve of U~Gem [see, e.g., fig.~14 of Bruch (2021)], with
a not very deep eclipse, preceeded
by a strongly expressed orbital hump. After eclipse
a much fainter secondary hump appears. The
similarity between V447~Lyr and U~Gem also extends to their outburst
characteristics: A small number of shorter and slightly fainter outbursts
are interspersed between longer and brighter ones 
[see fig.~2 of Ramsay et al.\ (2012) and the AAVSO long term light curve of 
U~Gem]. Even the orbital periods only differ by just half an hour.

Ramsay et al. (2016) already studied the waveform of the orbital variations of
KIC~9202990. Therefore, it is not surprising that the average (black
in Figure 1) is quite similar to their fig.~4. 
Investigating the waveform as a function of the time interval between two
random epochs Ramsay et al.\ (2016) also detected that it depends
on the phase of the mini-outbursts in KIC~9202990 (their fig.~6). In
a simpler approach I separately constructed 
phase folded light curves for the brighter and the fainter intervals of
KIC~9202990 (separated by the average flux). The 
results are shown in red (bright phases) and
blue (faint phases) in Figure~1. Five features are immediately 
obvious: (i) the amplitude increases with increasing
brightness, (ii) the width of the minimum, which appears as a shallow 
eclipse 
of an extended light source, is reduced when the brightness is low, (iii)
the minimum occurs slightly earlier during the faint, and later
during the bright phases, (iv) during the faint phases a secondary minimum
develops in the waveform which is
absent during the bright phases, and (v) the orbital hump is structured,
exhibiting a slightly rising plateau close to maximum when KIC~9202990 is faint
and developping an additional spike upon the hump when it is bright. On the
whole the various elements of the waveform are quite similar to those of
V447~Lyr (or U~Gem), but the minimum is shallower and the
secondary hump only stands out during the fainter phases.
These details may tell us a lot about the structure of KIC~9202990, but 
a more thorough analysis is beyond the scope of this small contribution.

\section*{Acknowledgements}

This paper is based on data collected by the Kepler mission (funded by
the NASA Science Mission Directorate) and 
obtained from the MAST data archive at the STScI operated by AURA
under contract NAS 5–26555.

\section*{References}

\begin{description}
\parskip-0.5ex

\item Bruch A., 2021, MNRAS, 503, 953
\item Kato T., Osaki Y., 2013, PASJ, 65, 97
\item Lomb N.T., 1976, Ap\&SS, 39, 447
\item Osaki, Y., Kato, T. 2013a, PASJ 65, 50
\item Osaki, Y., Kato, T. 2013b, PASJ 65, 95
\item Ramsay G., Cannizzo J.K., Howell S.B., et al., 2012, MNRAS, 425, 1479
\item Ramsay G., Hakala P., Wood M.A., 2016, MNRAS, 455, 2772
\item Savitzky A., Golay M.J.E., 1964, Analytical Chemistry, 36, 1627
\item Scargle J.D., 1982, ApJ, 263, 853
\item Schwarzenberg-Czerny A., 1989, MNRAS, 241, 153
\item Thorstensen J.R., Taylor C.J., 1997, PASP, 109, 1359
\end{description}

\end{document}